\documentclass[11pt]{article}
\usepackage{hyperref,xcolor}
\usepackage{cite}
\usepackage{footnote}
\makesavenoteenv{minipage}

\newcommand{\mc}{\mathcal}
\newcommand{\p}{$\mathcal{PT}$ \ }

\renewcommand{\title}[1]{%
    \bigskip%
    \begin{center}%
    \Large\bf #1%
    \end{center}%
    \vskip .2in}

\renewcommand{\author}[1]{%
    {\begin{center}
    #1
    \end{center}}}
\newcommand{\address}[1]{\vspace{-1.7em}\vspace{0pt}
    {\begin{center}
    \it #1
    \end{center}}}

\begin{document}

\begin{titlepage}
\title{Treating Ostrogradski instability for Gallilean invariant Chern Simon's model via \p symmetry }

\author
{
Biswajit Paul  $\,^{\rm a,b}$,
Himangshu Dhar $\,^{\rm a,c}$,
Mangobinda Chowdhury $\,^{\rm a, d}$,
Biswajit Saha    $\,^{\rm a, e}$}
\address{$^{\rm a}$ National Institute of Technology Agartala \\
 Jirania, Tripura -799 055, India }

\footnote{
{$^{\rm b}$\tt biswajit.thep@gmail.com,}
{$^{\rm c}$\tt himangshu171@gmail.com,}
{$^{\rm d}$\tt mangobinda4u@gmail.com,}
{$^{\rm e}$\tt biswajit.physics@gmail.com}}
\begin{abstract}

The Ostrogradski ghost problem that appears in higher derivative theories containing constraints  has been considered here. Specifically we have considered  systems where only the second class constraints appear. For these kind of systems, it is not possible to gauge away the linear momenta that cause the instability. To solve this issue, we have considered the PT symmetric aspects of the theory. As an example we have considered the Galilean invariant Chern Simons model in $2+1$ D which is a purely second class system. By solving the constraints, in the reduced phasespace, we have derived the \p similarity transformed Hamiltonian and putting conditions on  we found that the final form of the Hamiltonian is free from any linear momenta and  bounded from below. 
\end{abstract}
\end{titlepage} 
	\section{Introduction}
Higher derivative(HD) theories are important when one considers renormalisation \cite{stelle}, regularisation  and problems like ultraviolet divergences. By higher derivative we refer to those theories where the Lagrangian explicitly contains higher time derivatives of the field variables. Sometimes these HD theories are condidered as some correction terms or perturbative terms. In modern day theoretical physics, application of HD theories are wide e.g. cosmology     \cite{neupane, nojiri4}, supersymmetry   \cite{Iliopoulos, Gama}, noncommutativive theory      \cite{clz}, gravitation \cite{accioly, soti, gullu, ohta} etc.  Despite their wide range of usage these theories are plagued by a crucial issue called the `Ostrogradski instability'. HD theories are in general degenerate in nature and for this reason, while quantising, one should  very carefully choose the proper phasespace. It has been seen that Hamiltonian of the HD theories contain terms linear in some of the momenta. Now, when quanisation is done in the proper phase space, after removing the constraints, these linear momenta terms give rise to states with negative norm. It is because the Hamiltonian is not bounded from below and consequently the quantised theory becomes unstable. This instability is very serious and an open problem.
	 
	 The concept of $\mc{PT}$ quantum mechanics was introduced by Bender \cite{bender1998}. It was mainly introduced to deal with the non Hermitian Hamiltonians that initially appeared in cases like the quantum system of hard spheres \cite{wu}, Reggeon Field theory\cite{brower} and Lee Yang edge singularity \cite{cardy, zamolochikov}.  $\mc{PT}$symmetric Hamiltonians, though complex and  non Hermitian in nature, can have eigen value that are real \cite{bender1998}. Since then a series of investigations began on various properties of the $\mc{PT}$ symmetric Hamiltonians. It includes various discipline like nonlinear lattices \cite{abdullev}, waveguides with small holes \cite{borisov}, graphene nanocarbons \cite{fagoti}, quantum dot \cite{zhang2}, electrodynamics \cite{milton}, quantum field theory \cite{yang, goksel} etc.  As soon as   introduction of the theory there is also various experimental confirmation of the \p symmetry \cite{ruter, zhang}. Despite non-Hermitian nature of the \p  Hamiltonians  they were shown to have real spectra \cite{bender1998}.  
	
	Usually, the Euler-Lagrange equations of motion for a higher derivative theory are calculated using Ostrogradski formulation \cite{ostro}. In this method, all the momenta corresponding to the higher derivatives are defined in a nontrivial way. However,  recently it has been shown that this nontrivial Hamiltonian formulation, for some models, leads to an incorrect calculation of the phase space \cite{bmp1} and hence gives a misinterpretation of the quantum states. To bypass this issue  HD models were, rather, treated as  first order systems via redefintion of the fields. According to this first order formulation,  momenta definitions are usual. Surprisingly, in both the cases, Ostrogradski method as well as the first order formulation, the canonical Hamltonian contain terms linear in some of the momenta. Classically, these linear momenta can access both the positive and negative axes of the phasespace and the Hamiltonian as a  result of this  become unbounded.  After quantisation  these linear momenta terms  give rise to negative norm states which are  called  the ghost states \cite{woodard}. All these facts are well known for HD theories. The problem becomes more complicated if the systems are nondegenerate. According to Ostrogradski theorem, all the nondegerate systems  contain the ghost states. As a solution to this problem,  there exist few but  successful model dependent procedures adopted by different authors \cite{chen, klein}. HD models with constraints  may have first class and second class constraints or both. In \cite{bp}, the issue of ghost problem was addressed and solved for a system having first class constraints but this procedure lacks any scope  for  the systems only with second class constraints. 
	
	In this paper, we have considered the case of a purely second class systems where one cannot gauge away the constraints to remove the linear unwanted momenta. Actually for the second class systems, the reduced phasespace is obtained by solving the constraints and treating them  just as identity between the phasespace variables. Even in the reduced phase space, the linear momenta remain in the Hamiltonian and consequently the system becomes unbounded along that particular momenta. This problem of irreducibility of the linear momenta has been addressed in the present paper.  For that we have considered the approach of Bender et al.\cite{bender1} where applying a similarity transformation on the $\mc{PT}$ symmetric HD Hamiltonian they have obtained a bounded from below Hamiltonian. In the present paper we have generalized this approach for the second class systems.  
	
	 To implement the method discussed above we have considered a model, namely the Galilian invariant Chern Simon's model in 2+1 dimension, which has only second class constraints. First introduced by Lukiersky \cite{lukiersky}, this Chern Simon's model in 2+1 dimensions have already been explored by  many authors which includes interesting results relating Noncommutative geometry \cite{jackiw, hovarthy}, Berry phase \cite{duval}, Hall effect \cite{duval2}, Newton-Hooke symmetry \cite{alvarez}, twistors \cite{fedoruk}, anyon etc. The model has the symmetries of the Gllilie group i.e. time translation, space translation, boost and rotation \cite{lukiersky}. This type of Chern Simon's models have a central charge, in this case it is $m$, and an additional central extention can provided only for the $D=2+1$ which form the exotic Gallilie algebra \cite{fedoruk2}. In fact the model in \cite{lukiersky} was shown to have same dynamics as a charged non-relativistic planar particle in external homogeneous electric and magnetic field\cite{olmo}. 
	 
	 The organisation of the paper is as follows. In Sec II we shall consider the HD theory and its first order formulation. Sec III deals with the \p symmetric nature of these HD theories. In Sec IV we shall describe the 2+1 Gallilean invariant Chern-Simons model and its constraint analysis by  applying the Dirac method. In Sec V we shall perform the \p symmetric transformations to obtain the  bounded from below Hamiltonian of the model. Finally, we conclude with Sec VI.

	\section{Higher derivative models}
    A typical  higher derivative Lagrangian is written as 
	\begin{equation}
	L = L(Q, \dot{Q}, \ddot{Q}...Q^{(n)}).
	\end{equation}
	Here 'dot' represents time derivative of the fields $Q$. To avoid the Ostrogradski's way of Hamiltonian  formulation we redefine the configuration space by incorporating new variables as  $Q=q, \dot{Q}=q_1, \ddot{Q}=q_2 ......  Q^{(n-1)}=q_n$. So the new Lagrangian is
	\begin{equation}
	L' = L(q, q_1, q_2...q_n) + \sum_{i=1,n}\lambda_i\zeta_i. \label{auxlag1}
	\end{equation}
Here the $\lambda_n$'s are Lagrange multipliers incorporated to account the constraints
\begin{equation}
\zeta_i = q_i -\dot{q}_{i-1} 
\end{equation} 
in the new  configuration space.  It is evident that the Euler Lagrange equation of motion for the Lagrangian $(\ref{auxlag1})$ now will be first order. The momenta can be found as
\begin{eqnarray}
p_i &=& \frac{\partial L'}{\partial \dot{q}_i}, \\ 
p_{\lambda_i} &=& 0.
\end{eqnarray}
Not all the momenta defined above are invertible for a constraint system. Hence these momenta will, in general, give rise to the primary constraints at this stage. The  Hamiltonian  can be written as
\begin{eqnarray}
H _{can} = \sum_{i=1,n} \Big( p_i\dot{q}_i + p_{\lambda_i}\dot{\lambda}_i \Big) - L'. \label{can_ham}
\end{eqnarray}
It is to be noted that at this level the canonical Hamiltonian when simplified will  contain terms  like $p_{i-1}q_i$. These linear momenta will populate the positive as well as negative regions of the phasespace and hence will give rise to sates in quantised theory that have  negative norms. The interesting fact is that this is a model independent outcome for all HD theories. 

The total Hamiltonian of the theory is
\begin{equation}
H_T = H_{can} + \Lambda_i \Phi_i.
\end{equation}
		Here $\Lambda_i$'s are Lagrange multipliers which act as coefficients linear to the primary constraints. Next we find out the time evolution of the primary constraints. Depending on the systems  two scenarios can appear
		
\begin{enumerate}
\item[i.] PBs of the primary constraints with the Hamiltonian are equal to some function of the phase space variables but they do not include any of the Lagrange multiplier $\Lambda_i$'s. In this case  they are treated as secondary constraints.

\item[ii.] If the PBs of the primary constraints with the Hamiltonian involve Lagrange multiplier $\Lambda_i$, then equating them to zero we can find out the Lagrange multipliers. 		
\end{enumerate}

 In order to find out all the constraints of the theory at the secondary, tertiary levels  the iteration process goes on . We may group all the constraints all these constraints as first class and second class based on the nature of the PBs among themselves. In the present paper we are interested in systems with only second class systems. Evidently, in this case, all the Lagrange multipliers are solved and can be used in the total Hamiltonian. In the reduced phase space, we can get rid of all the second class constraints by treating them as indentities. One can expect that these constraints might involve momenta like $p_{1i}$ and in the reduced phase space the Hamiltonian is free from these unwanted  linear momenta. Unfortunately, such is not the scenario in most of the cases. Rather, the Hamiltonian till is plagued with the linear momenta and is unbounded from below. A mere reduction of the phase space by applying the Dirac procedure is no help for us because the  system has only  second class constraints. 
\section{PT symmetries:}
In this section we point out only the basic features of  $\mathcal{PT}-$symmetric quantum mechanics required to reach our goal of finding out a bounded from below Hamiltonian for the Higher derivative models only with second class constraints.

  Consider a linear operator $\mathcal{P}$ that effects any other operator,  some function of position and momentum, through spatial reflection. Effect of  $\mathcal{{P}}$ on the basic operators like space $\hat{x}$ and momentum  $\hat{p}$ is:
  \begin{eqnarray}
\mathcal{P} \hat{x}=-\hat{x},\\
\mathcal{P} \hat{p} =-\hat{p} .
\label{Ptrans}  \end{eqnarray}  
Also consider another operator which is antilinear and effects the position and momentum as  
\begin{eqnarray}
\mathcal{T} \hat{p}=-\hat{p}, \\
\mathcal{T} \hat{x}= \hat{x}.
\label{Ttrans}
\end{eqnarray} 
So, it is evident that a combination of the \p operation left the momentum operator $\hat{p}$ unchanged but the  position operator $\hat{x}$ changes sign. States which are eigenstates of the Hamiltonian are also simultaneously eigenstates of \p.

 We know that the Hamiltonian of a theory gives the energy eigen values of a system. If the Hamiltonian is real and symmetric  consequently the  energy eigen values are also real.  Although there is a stronger condition for this reality of the energy spectrum and it is called   Hermiticity of the Hamiltonian. For real spectrum of the theory in \cite{bender2005} C. M. Bender showed that  Hermiticity  of the Hamiltonian as sufficient but not necessary condition.  In usual theories it is expected that the Hamiltonian is  bounded from below but  the Higher derivative theories are  usually devoid of this boundedness due to existence of the Ostrogradski ghosts. In the  general form as the equation (\ref{can_ham})  contain linear momenta terms like $p_{1i}q_{2i}$, it is not bounded from below. This linear momenta, in coordinate space, will make the Hamiltonian a  complex operator and consequently will become nonHermitian $H \ne H^\dagger$. To overcome this condition Bender et al proposed the concept that if the Hamiltonian is complex and if it is not Hermitian then for real energy spectrum it should be \p symmetric i.e. $H = H^{\mathcal{PT}}$. To dig more facts and outcome of \p theory the reader may refer to \cite{bender2005,bender2004}.

  One might be curious about the state vectors and how the Hilbert space is defined\cite{bender2002}. The state vectors for these complex Hamiltonians give rise to negative norm states while for a viable physical theory it is required that the norm of the state vectors must be positive. For these type of complex Hamiltonians, there exist a previously unnoticed symmetry operator denoted by $\mathcal{C}$  with the properties 
 \begin{equation}
 \mc{C}^2=1, \ \ \ \ [\mc{C},\mc{P}\mc{T}]=0, \ \ \ \ [\mc{C},H]=0. \label{identity}
 \end{equation}
 In \p  theory the operator $\mc{C}$ is defined in such a way so that the inner product  
 \begin{equation}
 <\psi|\chi> = \int dx [\mathcal{CPT}\psi(x)] \chi(x)
 \end{equation}
 is positive definite.  This $\mathcal{C}$ operator has similarities with the charge conjugation operator. $\mc{C}$ can be written as \cite{bender2003}
 \begin{equation}
 \mc{C} = e^Q \mc{P}, \label{c_operator}
 \end{equation}
 where $Q$ is a real function of the dynamical variable or the phase space variables. 
	\section{The 2+1 Chern Simon's model : first order treatment}
As shown in \cite{lukiersky} we consider the Gallilean Invariant model in $D=2+1$ dimensions 
\begin{eqnarray}
L = \frac{m\dot{x}_i^2}{2} - k \epsilon_{ij}\dot{x}_i \ddot{x}_j
\end{eqnarray}
where $k$ has a physical dimension of $[M][T]$. We convert the Lagrangian into a first order form by redefining the field variables as 

\begin{eqnarray}
q_{1i} = x_i, \\
q_{2i} = \dot{x}_i. 
\end{eqnarray}

So in terms of these variables the Lagrangian becomes 
 \begin{eqnarray}
 L = \frac{mq_{2i}^2}{2} - k \epsilon_{ij}q_{2i} \dot{q_{2j}}+\lambda_{i}(\dot{q}_{1i}-q_{2i}).
  \end{eqnarray}

Corresponding to this Lagrangian, the momenta are given by
\begin{eqnarray}
\nonumber 
p_{1i} &= &\frac{\partial L}{\partial \dot{q}_{1i}} = \lambda_{1i}, \\
p_{2i}& =& \epsilon_{ij}q_{2j}, \\ 
p_{\lambda i}& =& 0.
\end{eqnarray}
Hence we get the primary constraints which are
\begin{eqnarray}
\Phi_{i} &=& p_{1i} - \lambda_{i} \approx 0, \\
\Psi_{i} &=& p_{2i} - k \epsilon_{ij}q_{2j} \approx 0, \\
\Xi_{i} &=& p_{\lambda i} \approx 0.
\end{eqnarray}
Poission brackets among these primary constraints are
\begin{eqnarray}
\{\Phi_{i},\Psi_{j} \} &=& 0, \\
\{\Psi_{i},\Psi_{j} \} &=& -2k\epsilon_{ij},\\
\{ \Phi_{i}, \Xi_j\} &=& -\delta_{ij}, \\
\{\Psi_i, \Xi_j\} &=& 0.
\end{eqnarray}
The canonical Hamiltonian is 
\begin{eqnarray}
H_{can} &=& p_{1i}\dot{q}_{1i}+p_{2i}\dot{q}_{2i}+p_{\lambda i}\dot{\lambda}_{i} - L \nonumber \\ 
&=& -\frac{m}{2}q_{2i}^2 + \lambda_iq_{2i}
\end{eqnarray}
Whereas the total Hamiltonian is 
\begin{eqnarray}
H_T = H_{can} + \Lambda_{1i}\Phi_{i}+ \Lambda_{2i}\Psi_{i}+ \Lambda_{3i}\Xi_{i}.
\label{totH}
\end{eqnarray}
At this level we want to find out the various Lagrange multipliers by computing the Poission brackets of the primary constraints with the total Hamiltonian and equating them to zero, which gives
\begin{eqnarray}
\Lambda_{3i}&=&0, \\ 
\Lambda_{2i} &=& \frac{1}{2k}(mq_{2j}+\lambda_j)\epsilon_{ji},\\
\Lambda_{1i} &=& -q_{2i}.  
\end{eqnarray} 
Thus we see that there is no generation of new constraints. The chain of constraints stops here with the second class constraints. However we can remove these constraints as they are second class in nature. The constraints $\Phi_i$ and $\Xi_i$ can be removed by simply putting them to be zero. Corresponding Dirac brackets between the phase space variables remain unchanged and hence all the Poission brackets those were computed earlier remain same under these Dirac brackets. Now we are left with the second class constraints $\Psi_i$. For their removal and  construction of the Dirac brackets we consider the matrix 
\begin{equation}
\Delta_{ij} = \{ \Psi_i, \Psi_j\} = -2k\epsilon_{ij} 
\end{equation} 
and it's inverse 
\begin{equation}
\Delta_{ij}^{-1} =  \frac{1}{2k}\epsilon_{ij}.
\end{equation} 
Removing the second class constraints we get 
\begin{equation}
H_T=H_{can}.
\end{equation}
We observe that  the Hamiltonian till contains the term linear in the momenta which actually gives rise to ghost states i.e. the negative norm states. Although it is a purely second class theory and for that reason we don't have any way to get rid of the ghost fields by just incorporating new constraints in the form of gauge conditions. 

 In the next section we shall try to see if the theory is $\mathcal{P}\mathcal{T}$ symmetric or not. This will enable us to apply the treatment of \cite{bender1} to form a ghost free Hamiltonian. 
    
\section{$\mathcal{P}\mathcal{T}$ symmetry of the Gallilian invariant Chern Simon model:}
In this section we shall try to fix this problem of removing the ghost fields by considering the $\mathcal{P}\mathcal{T}$ version of the model. After solving the second class constraints the total Hamiltonian (\ref{totH}), in the reduced phase space, becomes  
\begin{equation}
H = -\frac{m}{2}q_{2i}^2 + p_{1i}q_{2i}. 
\label{hamQ}
\end{equation}
This Hamiltonian contains a term $p_{1i}q_{2i}$ which is linear in  momenta. The Hamiltonian thus is not bounded from below. We  consider the isospectral(similarity) transformation as 
\begin{eqnarray}
q_{1i} = iz_i, \\ 
p_{1i} = -ip_{zi}. \label{isospectral}
\end{eqnarray}
Replacing these in the Hamiltonian we find out the new Hamiltonian which is :
\begin{equation}
\tilde{H} = -\frac{m}{2}q_{2i}^2 -       ip_{zi}q_{2i}. \label{hamPT}
\end{equation}
Now, if we apply the transformations of (\ref{Ptrans}, \ref{Ttrans}) it can easily be seen that the Hamiltonian obtained above is $\mathcal{P}\mathcal{T}$ symmetric. This Hamiltonian, however, is neither bounded from below nor real. Also, this Hamiltonian is not even Dirac hermitian. To check if the removal of these ghost fields are really possible or not, we shall analyze as the following. We consider the  usual transformation of the fields as
\begin{eqnarray}
\nonumber 
\hat{\mathcal{P}}z_i &=& z_i, \\
\nonumber 
\hat{\mathcal{P}}p_{zi} &=& p_{zi}, \\
\nonumber 
\hat{\mathcal{P}}q_{2i} &=& -q_{2i}, \\
\nonumber 
\hat{\mathcal{P}}p_{2i} &=& -p_{2i}. \\
\end{eqnarray} 
 Whereas, under the time reversal operator the transformations are 

\begin{eqnarray}
\nonumber 
\hat{\mathcal{T}}z_i &=& -z_i, \\
\nonumber 
\hat{\mathcal{T}}p_{zi} &=& p_{zi}, \\
\nonumber 
\hat{\mathcal{T}}q_{2i} &=& q_{2i}, \\
\nonumber 
\hat{\mathcal{T}}p_{2i} &=& -p_{2i}. \\
\end{eqnarray} 
Effect of the combined $\mathcal{P}\mathcal{T}$ transformation $\tilde{H}$ leaves invariant
\begin{equation}
\mathcal{P}\mathcal{T} \tilde{H}(\mathcal{P}\mathcal{T})^{-1} = \tilde{H}.
\end{equation}
Although the  Hamiltonian (\ref{hamPT}) is not Dirac hermitian but it is  $\mathcal{P}\mathcal{T} $ symmetric. Reason for considering $\mathcal{P}\mathcal{T} $ symmetric Hamiltonian is that they give real spectrum despite unitarity is violated.

 To find the $\mathcal{C}$ operator as defined in (\ref{c_operator}) we first find out $Q$. Although $Q$ can be obtained perturbatively but we adopt the definition given in \cite{bender1} which is a bilinear function of the phase space variables, which, for the present model of the paper with the phase space $\{ q_{1i}, p_{1i}, q_{2i}, p_{2i} \}$ becomes 
\begin{equation}
Q = \alpha p_{zi}p_{2i} + \beta z_{i}q_{2i}.
\end{equation}
Where $\alpha, \beta$ are some parameters yet to be determined. 
\subsection{Finding $\alpha$ , $\beta$:}
To find the parameters $\alpha$ and $\beta$ we recall the relations in (\ref{identity}). The first two relations give an idea aout  the properties of the $\mathcal{C}$ operator (see(\cite{bender2005} for detail explanation)). The third relation is interesting as one can see that the $\mathcal{C}$ commutes with the Hamiltonian. Using this identity and nonHermitian nature of the Hamiltonian we can easily write 
	\begin{eqnarray}
	e^{-Q}He^{Q}=H_0-H_1. \label{H_pt}
	\end{eqnarray}
The Hamiltonian as always, for unbroken $\mathcal{PT}$ theories, can be decomposed into two parts with $H_0$ being the usual kinetic part which is Hermitian by construction  and the $H_1$ contains the non-Hermitian term. 

Now we  calculate the similarity transformations of the phase space variables, which are
\begin{eqnarray}
\nonumber
e^{-Q} z_{i}e^{Q} &=&  z_i C + DSp_{2i},\\
e^{-Q} p_{zi}e^{Q }&=& p_{zi}C + \frac{S}{D}q_{2i},\\ 
e^{-Q} q_{2i}e^{Q} &=& q_{2i} C + DSp_{2i},  \\
e^{-Q} p_{2i}e^{Q} &=& p_{2i}C + \frac{S}{D}z_i .
\end{eqnarray}
Where we have taken $D= \sqrt{\frac{\alpha}{\beta}}$ , $S = sinh\sqrt{\alpha \beta}$, $C=cosh\sqrt{\alpha \beta}$. Using these transformation rules one can easily calculate the transformations for the functions. The Hamiltonian transforms as 
\begin{eqnarray}
\nonumber
e^Q \tilde{H} e^{-Q} &=& -\frac{m}{2}\Big( q_{2i}^2 cosh2\sqrt{\alpha\beta}-\frac{p_{zi}^2}{2}D^2 (cosh2\sqrt{\alpha\beta}-1) \Big)+\Big(\frac{q_{2i}^2}{D} sinh2\sqrt{\alpha\beta} -\frac{p_{zi}^2}{2} Dsinh2\sqrt{\alpha\beta} \Big) \\ &&  -i  p_{zi}q_{2i} \Big(\frac{m}{2}Dsinh2\sqrt{\alpha\beta}  +cosh2\sqrt{\alpha\beta}\Big) . \label{H_PT}
 \label{transH}
\end{eqnarray}
Putting this expression in (\ref{H_pt}) and using the expressions of $H_0$ and $H_1$ we get an identity. Comparing the imaginary on oth the sides we get a simplified form 

\begin{equation}
cosh\sqrt{\alpha\beta} = -\frac{m}{2} \sqrt{\frac{\alpha}{\beta}} sinh\sqrt{\alpha\beta}. \label{ab_relation}
\end{equation} 

This is a very important relation we have otained between $\alpha$ and $\beta$.

\subsection{The bounded from below Hamiltonian:}
It is clear that the linear momenta terms are till present in the equation (\ref{H_PT}). These terms can be removed if either they are constraints of the system or the coefficients are such that they yield to zero. Dirac constraint analysis performed in the previous section suggest that in the present system $p_{zi}q_{2i}$ do not form any constraint. On the other hand, since it is a purely second class system we can not remove them by introducing these terms as gauge conditions.
Contrary to this approach of Dirac constraint analysis, since this is a $\mathcal{PT}$ symmetric system, we can apply similarity transformation to (\ref{hamQ}) and using  (\ref{ab_relation}) we get 
\begin{equation}
\tilde{H}'=e^{-Q/2}He^{Q/2}= q_{2i}^2 \Big(\frac{m^2}{4}DS+\frac{S}{D} \Big) + \frac{p_{zi}^2}{2}\Big(\frac{m}{2}D^2(C-1)+\frac{2}{m}C\Big)  . \label{transH1}
\end{equation}
 It is eident that all the terms inoling phase space variables are in the squared form and consequently the above transformed Hamiltonian is bounded from below. Thus the $\mathcal{PT}$ transformations helped us to to get rid of the unwanted linear momenta that can give rise to negative norm states. 
The eigen states $|\tilde{\psi}>$ of the Hamiltonian (\ref{transH1}) have a positive inner product and is normalised $<\tilde{\psi}|\tilde{\psi}>=1$ in the Dirac sense. This is gurranteed because   the states of the original Hamiltonian (\ref{hamPT}) are connected via $|\tilde{\psi}> = e^{-Q/2}|\psi>$. In \p theory these eigen states are normalised as 
\begin{equation}
<\psi_n|e^{-Q}|\psi_m> = \delta (m,n), \sum |\psi_n><\psi_n|e^{-Q} = 1.
\end{equation}
              
Thus for the second class systems the above algorithm may be useful for other models also. Provided that after  similarity transformation (\ref{isospectral}), the Hamiltonian should be \p symmetric. For those models the unboundedness may be just an artifact which can be removed by taking into account a more physical symmetry viz. the \p symmetry.
\section{Conclusions}
 Higher derivative theories have been seen in active use in various branches of physics including gravity, string theory, condensed matter physics etc. One might argue the validity of these theories as there are adequate reasons to question on and one of them is the Ostrogradski ghost problem \cite{ostro, woodard}. To remove these unphysical sectors from the theory, earlier attempts were purely system dependent \cite{chen,bp}. In general, these HD models are always contain constraint  which includes both first class and second class constraints. The problem for solving the first class systems are proposed in \cite{bp} but the method can not deal with systems with only second class constraints. This issue was addressed here and a way to solve  by considering the \p symmetries of the system has shown.

 We have converted the HD theory into a first order theory by redefining the field variables and incorporating appropriate constraints at the Lagrangian level. While performing the Hamiltonian analysis we have found out all the constraints in the theory. We have considered  the problem of systems only with second class constraints. As an example the Gallilean invariant Chern Simons model was considered in 2+1 D. These second class constraints were removed by treating them to be zero and consequently  replaced all the Poission brackets of the theory by appropriate Dirac brackets. The main concern of the theory was with the Hamiltonian which contained a term $p_{1i}q_{2i}$ . This term, linear in momenta, gives rise to negative norm state when one goes for quantisation of the theory. It is evident from the constraint structure that none of the constraints, which are second class indeed, contain the momenta and hence we have no way to reduce this to get the reduced phase space. To solve this issue of the linear momenta we have considered \p symmetirc aspects of the model. Due to  existence of the linear momenta, after performing an isospectral similarity transformation, the Hamiltonian becomes imaginary. We noticed that although the Hamiltonian now became nonHermitian but it is \p symmetric. We calculated the \p transformations of the phase space variables. The \p transformation of the Hamiltonian is shown to be real and all the terms in the transformed Hamiltonian are bounded from below.   

 Though it is a higher derivative theory, finally we got the real and bounded from below Hamiltonian which is not usual for this class of theories. Despite its constraint structure the inherent \p symmetric nature of the theory was indeed a help to get rid of the linear momenta terms causing the instability of the quantum version. Now the Hamiltonian do not contain the Ostrogradski ghost term and is free from the instabilities. The method described above is applicable for systems where only second class constraints appear. It would be worth to investigate the applicability of this method on more complex and physically viable models.     
	 
\end{document}